# EIT-like transmission by interaction between multiple Bragg scattering and local plasmonic resonances

**Z Z Liu[1], Q Zhang[1] and J J Xiao[1],***

[1] College of Electronic and Information Engineering, Shenzhen Graduate School, Harbin Institute of Technology, Shenzhen 518055, China

*E-mail: eiexiao@hitsz.edu.cn

**Abstract**

We study the optical properties associated to both the polariton gap and the Bragg gap in periodic resonator–waveguide coupled system, based on the temporal coupled mode theory and the transfer matrix method. By the complex band and the transmission spectrum, it is feasible to tune the interaction between multiple Bragg scattering and local resonance, which may give rise to analogous phenomena of electromagnetically induced transparency (EIT). We further design a plasmonic slot waveguide side-coupled with local plasmonic resonators to demonstrate the EIT-like effects in the near-infrared band. Numerical calculations show that realistic amount of metal Joule loss may destroy the interference and the total absorption is enhanced in the transparency window due to the near zero group velocity of the guiding wave.

## 1. Introduction

It is well known that Bragg gap [1,2] and "polariton" gap or resonance gap [3,4] are two of the fundamental mechanisms to engineering photonic bands. Inside the gap range of frequencies, light cannot propagate through the structure. Basically, the Bragg gap originates from the interference of multiple Bragg scatterings in periodic microstructure, featuring it as an intrinsic consequence of long-range interference interaction. On the contrary, polariton gaps arise via a local atomic or structure resonance from either quantum transition or structurally confined interference, being generally not much related to the size scale of the sample. The Bragg gap is therefore much more sensitive than the polariton gap with respect to the sample size. From the transmission perspective, the sample size is more crucial in terms of stopping light by the Bragg gap than by the polariton gap.

While there have been lots of studies on the Bragg gap, as well as the polariton gap, respectively, their mutual interactions are attracting increasing concerns since they may yield interesting interference phenomena and unique properties in light propagation control. As a matter of fact, Fano resonance and electromagnetically induced transparency (EIT) can arise and be engineered in such a way, for example, in coupled ring-waveguide systems [5,6], photonic crystal based systems [7], and in metasurface arrays with simultaneous lattice resonance and local plasmonic resonance that can be tuned to be spectrally close [8]. Recently, Fano phenomenon by interference between localized magnetic plasmon resonance and Bragg resonance has been experimentally observed in metamaterials and was utilized to enhanced nonlinear third harmonic generation [9]. Therefore, phenomenological and straightforward approaches that can describe their interactions are highly desirable and shall be helpful for on-chip photonic design.

In this work, we present such a method that captures the essences of both the polariton gap and the Bragg gap simultaneously. The theory consists of two ingredients: (1) the temporal coupled mode theory (TCMT) [10-13] which describes the local resonance (either quantum or classical) in a very generic way. The TCMT gives a phenomenological transfer matrix for waveguide over the coupling "point" (interface) where the local resonance couples to the waveguide; (2) the transfer matrix method (TMM) [14,15], which describes the full transmission properties of the designed structure, as schematically shown in Fig. 1. By analyzing the optical propagation through a unit cell, we can get the field amplitude transfer matrices for both the unit cell and the total sample. Then, to characterize a periodic structure, we simply establish the dispersion relation by invoking the Bloch theorem, and the transmission spectrum of a finite-sized sample can be evaluated.

We would like to emphasis that phenomenological Hamiltonian approach [16] or spatial coupled mode theory [14] can also be utilized to describe similar Bragg-polariton interaction, these theories are not transparency as the present approach of TCMT plus TMM. Furthermore, our scheme is easy to adopt and flexible for different systems with cascaded side-coupled resonators.

## 2. Theoretical analysis by TCMT and transfer matrix method

The general scheme of the coupling system, which consists of a cascade of $n$ resonators side-coupled to a waveguide is shown in figure 1. The $i$ th resonator supports $m$ resonant modes and the normalized amplitudes of the resonances are denoted by a vector

$$\mathbf{a}_i = \begin{pmatrix} a_i^1 & a_i^2 & \cdots & a_i^m \end{pmatrix}^T, \qquad (1)$$

As shown in figure 1, $S_{\pm p}$ (p=1,2,3) represents the amplitudes of the incoming and outgoing waves at the "points" where the waveguide couples to the individual resonator. The transmission and reflection properties of such a system can be calculated by the TCMT. For an incoming wave at frequency $\omega$, the dynamics of $\mathbf{a}_i$ at the steady state can be expressed as [10,11]

$$\frac{d\mathbf{a}_i}{dt} = \left(-i\mathbf{\Omega} - \mathbf{\Xi}_0 - \mathbf{\Xi}_e\right)\mathbf{a}_i + \mathbf{\Gamma}^T \mathbf{S}_+, \qquad (2)$$

$$\mathbf{S}_- = \mathbf{C}_i \mathbf{S}_+ + \mathbf{B}_i \mathbf{a}_i, \qquad (3)$$

where $\mathbf{\Omega}$ ($\mathbf{\Xi}_e$) is a $m \times m$ matrix that represents resonant frequencies (coupling decay rate of the resonant modes), and $\mathbf{\Xi}_0$ is a $m \times m$ matrix that indicates the intrinsic decay rates. In Eqs. (2) and (3), $\mathbf{\Gamma}^T$ and $\mathbf{B}_i$ are the coupling matrices between the resonant modes and the incoming and the outgoing waves, respectively. $\mathbf{C}_i$ is the scattering matrix which describes the coupling between the incoming and the outgoing waves in the directive pathway. According to energy conversation and time-reversal symmetry, all the matrices are not independent [10].

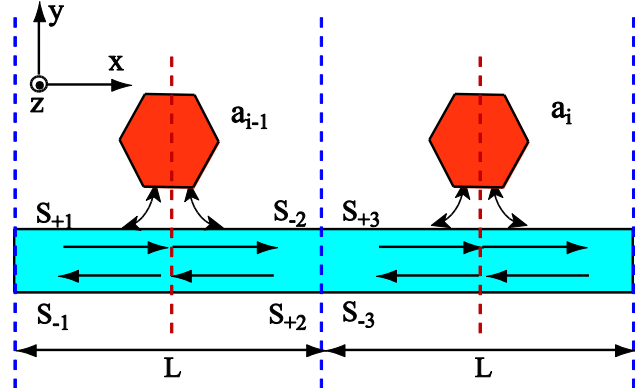

Figure 1. Schematic figure of periodic resonator-waveguide systems. The dashed blue line represents the boundary of a periodic unit.

In the structures we considered, only one channel is introduced (see figure 1). For incidence from the left port, we use the following matrices to describe the normalized wave amplitudes and their directive couplings

$$\mathbf{S}_- = \begin{bmatrix} S_{-1} \\ S_{-2} \end{bmatrix}, \ \mathbf{S}_+ = \begin{bmatrix} S_{+1} \\ 0 \end{bmatrix}, \ \mathbf{C}_i = \begin{pmatrix} 0 & 1 \\ 1 & 0 \end{pmatrix}.$$

For multi-channel structures, the system can also be described by Eqs. (2) and (3) and one just needs to adopt appropriate matrices for the respective cases.

Inside a unit cell, at the reference plane marked by dashed red lines in Fig. 1, there are established transmission and reflection coefficients $t$ and $r$, due to the presence of the resonator. In our method, we determine $t$ and $r$ by the means of TCMT, i.e., solutions from Eqs. (2) and (3). The field across the left and right sides of the $i$ th resonator reads

$$\begin{pmatrix} S_{-2} \\ S_{+2} \end{pmatrix} = \frac{1}{t}\begin{pmatrix} t^2 - r^2 & r \\ -r & 1 \end{pmatrix}\begin{pmatrix} S_{+1} \\ S_{-1} \end{pmatrix} = \mathbf{M}_1 \begin{pmatrix} S_{+1} \\ S_{-1} \end{pmatrix}. \qquad (4)$$

When the waves propagate in the waveguide over certain distance, a phase accumulation $\delta$ happens so that

$$\begin{pmatrix} S_{+3} \\ S_{-3} \end{pmatrix} = \begin{pmatrix} \exp(i\delta) & 0 \\ 0 & \exp(-i\delta) \end{pmatrix} \begin{pmatrix} S_{-2} \\ S_{+2} \end{pmatrix} = \mathbf{M}_2 \begin{pmatrix} S_{-2} \\ S_{+2} \end{pmatrix}, \tag{5}$$

Here $\delta = \beta L$ is the phase shift for the light propagating in the waveguide over a period length $L$, where $\beta$ is the propagation constant. It is straightforward that one can manipulate the position of the Bragg resonance in periodic structure by adjusting both $L$ and $\beta$. The total transfer matrix of a structure with $n$ cascaded unit can be written as $\mathbf{M} = (\mathbf{M}_2 \mathbf{M}_1)^n$ for the general configuration. The transmission of the structure is then defined by $T = |1/M_{22}|^2$.

To determine the dispersion relation of periodic structure (infinitely long), we apply the Bloch theorem [5,14,17] and require $\det|\mathbf{M} - \mathbf{I} e^{iK_B L}| = 0$, where $\mathbf{I}$ is the identity matrix, $K_B$ the Bloch wave number, and $L$ the period length of the structure along the propagation direction (see figure 1).

Next, we present several prototype cases with different resonant modes, constructed by cascaded resonant cavities. For simplicity, we neglect the intrinsic loss $\Xi_0$ in these discussions. The presence of it does not alter the physics. Firstly, in the case with one resonant mode in an individual resonator side-coupled to the incident channel, we can dictate $\mathbf{\Omega} = \omega_1$, $\Xi_e = \gamma_e$, $\mathbf{\Gamma}^T = i\sqrt{\gamma_e}$, $\mathbf{B}_i = i\sqrt{\gamma_e}$. Applying these to Eqs. (2) and (3), one immediately have the transmission and reflection coefficients

$$t = \frac{S_{-2}}{S_{+1}} = \frac{i(\omega_1 - \omega)}{i(\omega_1 - \omega) + \gamma_e}, \tag{6}$$

$$r = \frac{S_{-1}}{S_{+1}} = \frac{-\gamma_e}{i(\omega_1 - \omega) + \gamma_e}, \tag{7}$$

Figure 2 shows the complex band and transmission for such a single resonance with resonant frequency $\omega_1 = 2\omega_0$ and coupling constant $\gamma_e = 0.3\omega_0$. The upper panels in figure 2 are for the case where the Bragg resonance falls closely near the polariton resonance whereas the lower panels are for the case that they are spectrally separated. Obviously, for the former case [see Figs. 2(a)-2(c)], the interaction between the polariton gap and the Bragg gap gives rise to a mini-band which leads to high transmission in the otherwise stop-band, resembling the EIT phenomena. The effectiveness of our theory is manifested by tuning $\delta$ which governs the Bragg resonance position. When the Bragg resonance is tuned to $4\omega_0$ by modifying $L$, the Bragg resonance and the polariton resonance are fully separated [see Figs. 2(d)-2(f)]. Then we find classical line shapes of both Bragg gap and polariton gap in the complex band [see Figs. 2(d) and 2(e)]. The fact that the Bragg gap is more sensitive to the sample size is reflected by the relatively small value of the imaginary part of the Bloch wavevector $\text{Im}(K_B)$. Figure 2(a) and 2(d) both show that $\text{Im}(K_B)$ is remarkably larger near the polariton resonance than at a pure Bragg resonance. The transmission spectra of finite scale sample (with one unit cell and 8 periods) are calculated [see Figs. 2(c) and 2(f)]. Both are in good agreement with the dispersion relations. For samples with more period number, the transmission at the Bragg gap further decreases, down to nearly zero which is an inevitable result that Bragg gap results from the multiple scattering process.

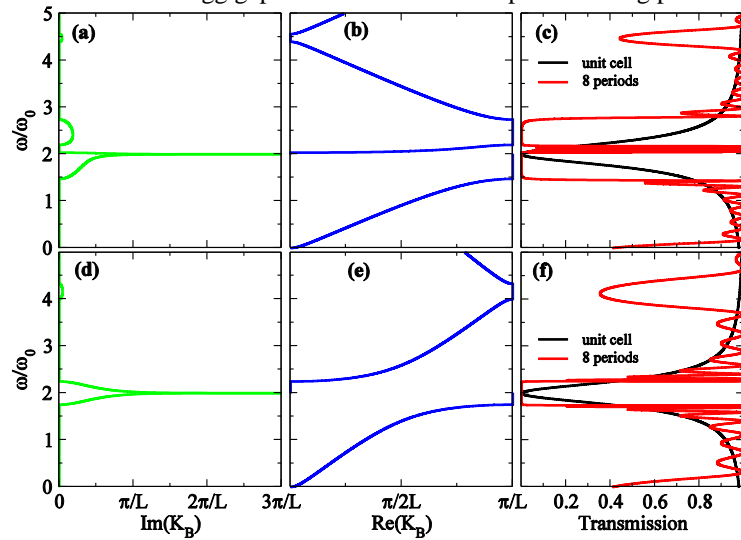

Figure 2. Complex band structure and transmission for the case of single resonance at $\omega_1 = 2\omega_0$ and coupling constant $\gamma_e = 0.3\omega_0$. (a)-(c) $\text{Im}(K_B)$ versus frequency, $\text{Re}(K_B)$ versus frequency, and the transmission for finite-sized sample of one unit cell (black line) and 8 unit cells (red line). The Bragg resonance is tuned to $2.2\omega_0$. (d)-(f) same as (a)-(c) but for Bragg resonance at $4\omega_0$.

Secondly, in the case of two coupling modes supported by two cascading resonators that are side-coupled to the incident channel, the TCMT contains the following matrices

$$\mathbf{\Omega} = \begin{pmatrix} \omega_1 & 0 \\ 0 & \omega_2 \end{pmatrix}, \ \mathbf{\Xi}_e = \begin{pmatrix} \gamma_e & i\mu \\ i\mu & 0 \end{pmatrix}, \ \mathbf{\Gamma}^T = \begin{pmatrix} i\sqrt{\gamma_e} & i\sqrt{\gamma_e} \\ 0 & 0 \end{pmatrix}, \ \mathbf{B}_i = \begin{pmatrix} i\sqrt{\gamma_e} & 0 \\ i\sqrt{\gamma_e} & 0 \end{pmatrix}.$$

Note that $\mu$ represents the coupling constant between the two cavities with intrinsic resonance frequencies $\omega_1$ and $\omega_2$, while $\gamma_e$ indicates the coupling strength between the cavity ($\omega_1$) and the waveguide channel. For this case, the transmission and reflection coefficients from Eqs. (2) and (3) read, respectively

$$t = \frac{S_{-2}}{S_{+1}} = 1 + \frac{-\gamma_e}{i(\omega_1 - \omega) + \gamma_e + \frac{\mu^2}{i(\omega_2 - \omega)}}, \quad (8)$$

$$r = \frac{S_{-1}}{S_{+1}} = \frac{-\gamma_e}{i(\omega_1 - \omega) + \gamma_e + \frac{\mu^2}{i(\omega_2 - \omega)}}. \quad (9)$$

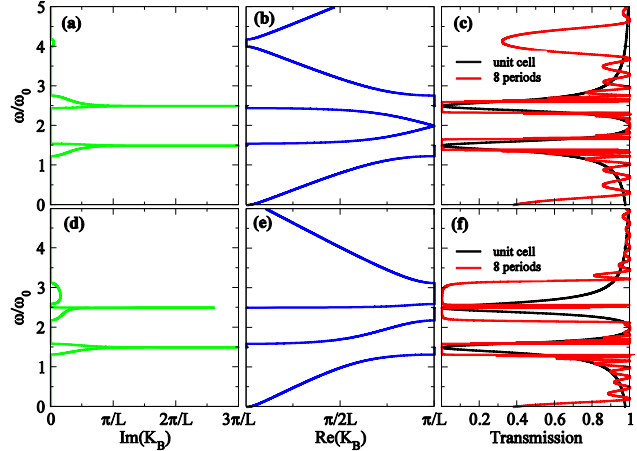

**Figure 3.** Complex band structure and transmission for the case of double resonances with resonant frequencies $\omega_1 = \omega_2 = 2\omega_0$ and coupling constant $\mu = 0.5\omega_0$, $\gamma_e = 0.3\omega_0$: (a-c) for Bragg resonance at $4\omega_0$ and (d-f) for Bragg resonance at $2.6\omega_0$.

In this case, we choose $\omega_1 = \omega_2 = 2\omega_0$, $\mu = 0.5\omega_0$ and $\gamma_e = 0.3\omega_0$. Then the resonant frequencies of the whole structure are determined by the hybrid mode resonances of the coupled double resonators [18]. The hybridization yields bonding and antibonding modes with frequencies [19,20]

$$\omega = \frac{(\omega_1 + \omega_2)}{2} \pm \frac{\sqrt{(\omega_1 - \omega_2)^2 + 4\mu^2}}{2} = (2 \pm 0.5)\omega_0, \quad (10)$$

i.e., polariton resonances. Figure 3 shows the complex band and the transmission spectrum for two different Bragg resonances. Similar to the single resonance case (figure 2), it is seen that the Bragg resonance and the polariton resonance can also be spectrally tuned to be overlapped or far-separated, yielding multiple mini-bands. Very interestingly, in this case, we can split the Bragg resonance gap to two connected polariton bands (upper panels of figure 3), yielding respective resonance gaps. Of course, Bragg resonance can also contain only one polariton resonance (lower panels in figure 3), simultaneously yielding EIT band and polariton gap. From figure 3(c) and 3(f) we can find that the two cascaded resonators form EIT-like transparency [21], while the band occurring in the upper Bragg gap in figure 3(f) results from the effective Fabry-Perot (FP) resonator, whose frequency is near the higher polariton resonance.

## 3. Design and numerical simulation of plamonic waveguide systems

The theoretical analysis above is very general and not limited to specific waveguide and local resonance nature. Here, as concrete examples, we design metal-dielectric-metal (MDM) plasmonic waveguides [22-24] and numerically verify the EIT-like behavior [21,25] by the finite element method (Comsol Multiphysics). Figure 4 shows two of the plasmonic slot devices with side-coupled Fabry-Perot (FP) resonators ($w = 50$ nm, $h = 110$ nm) filled with silica of $\varepsilon = 4.5$. The distance $d_1$ ($d_2$) is 20 nm (30 nm), which should be not so long as to ensure the effective coupling through evanescent wave. The medium in the slot waveguide is air, and the metallic cladding is silver with permittivity described by the well-known Drude model $\varepsilon_m(\omega) = \varepsilon_\infty - \omega_p^2/(\omega^2 + j\omega\Gamma)$ [26]. Here $\varepsilon_\infty$ is the dielectric constant at the infinite frequency, and $\Gamma$ and $\omega_p$ stand for the electron collision and bulk plasma frequencies, respectively. These parameters for silver can be set as $\varepsilon_\infty = 3.7$, $\omega_p = 9.1\,\text{eV}$, and $\Gamma = 0.018\,\text{eV}$.

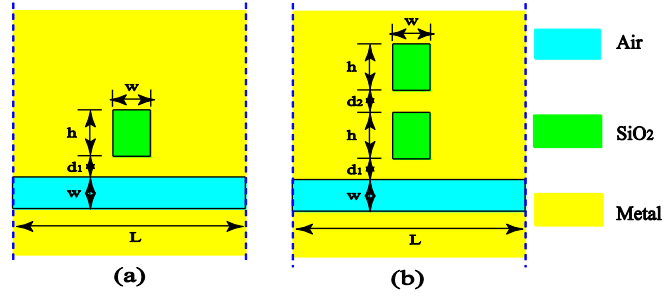

**Figure 4.** Schematic unit cell of the MDM waveguide-resonator system: (a) with one resonator, (b) with two cascaded resonators side-coupled to a waveguide.

Let us first consider the single-resonator case as shown in figure 4(a) at its fundamental TM mode resonance near $\lambda = 900$ nm. To figure out the interplay between Bragg resonance and polariton resonance, we set $\Gamma = 0\,\text{eV}$ and vary the period length $L$ to tune the position of Bragg resonance. This can be accomplished by $2n_{eff}L = m\lambda$ where the effective refractive index $n_{eff} = 1.385$ for $\lambda = 900$ nm. The integer $m = 1$ dictates the first Bragg gap position that appears near $\lambda = 900$ nm for $L = 326$ nm. Figure 5(a) shows the transmission for two different period $L = 326$ nm and $L = 500$ nm. Note that we have also used the unit cell result to obtain $\gamma_e$ (14.73 THz) in Eq. (7) and then get the theoretical transmission spectrum by the TMM. The results are also shown in Fig. 5(a) [labeled as TCMT-TMM]. It is seen that the numerical results and the theoretical ones agree quite well for both $L = 326$ nm and $L = 500$ nm. We can also observe that for the former case, the interaction between the polariton gap and the Bragg gap gives rise to an extremely narrow EIT-like transparency window [see the inset for zooming in], which is absent for the case of $L = 500$ nm which separates the Bragg gap from the polarition gap around $\lambda = 900$ nm. Unfortunately, since the group velocity $v_g = \partial\omega/\partial(\text{Re}(K_B))$ in the EIT window is nearly zero and lights stay in the structure for substantially long time, it is expected that absorption would dramatically affect the EIT phenomena. Indeed, figure 5(b) shows that for gradually increasing imaginary part of the silver dielectric function, the EIT-peak is suppressed and eventually disappears for the Drude model with $\Gamma = 0.018\,\text{eV}$. The total absorption $1 - T - R$, however, is enhanced by the EIT-like effect (figure not shown).

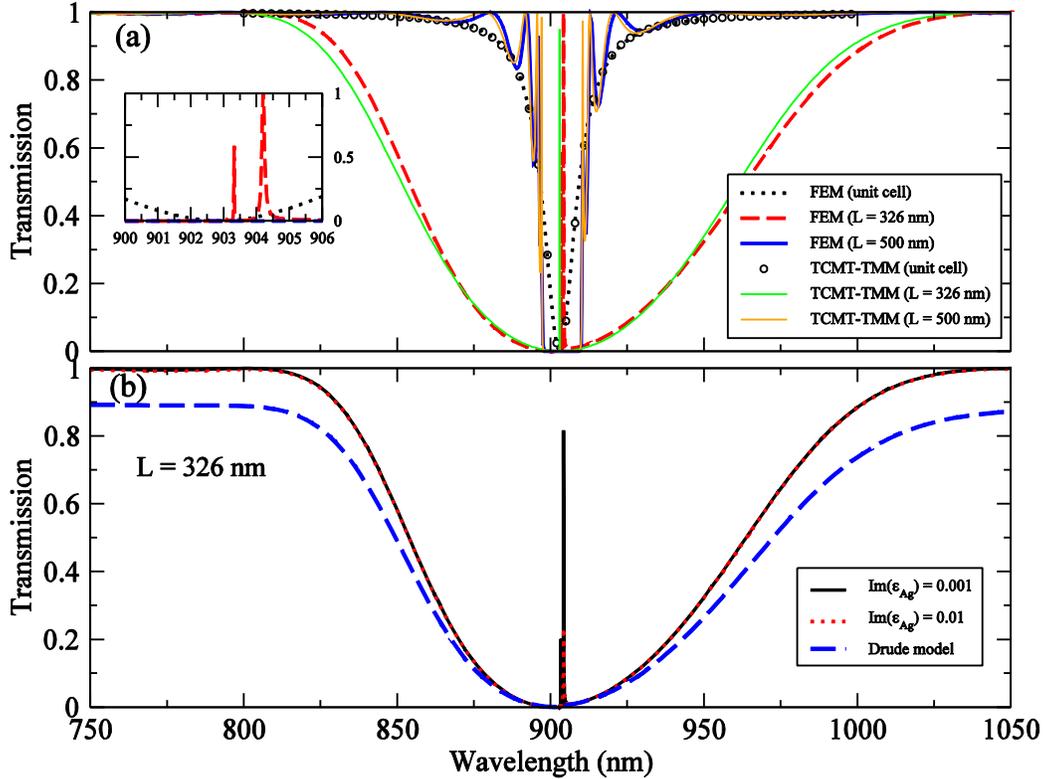

Figure 5. (a) Ful-wave numerical transmission spectrum (FEM) of periodic MDM structure in figure 4(a) for a unit cell (dotted black), period length $L = 326$ nm (dashed red) and $L = 500$ nm (full blue). Also the corresponding TCMT-TMM results of unit cell (circle black), $L = 326$ (full green) and $L = 500$ nm (full orange) are shown. The results by FEM and TCMT-TMM are in good agreement. Absorption loss is neglected. (b) Numerical transmission for different silver loss when $L = 326$ nm. All the period number is 8.

Figure 6 shows the normalized amplitude |H| along the waveguide for specific wavelengths for the cases in figure 5(a). Insets in each panel are the steady-state magnetic field distributions. Figures 6(a) and 6(c) demonstrate the typical Bragg gap stopping of wave propagation as the field penetrates to certain depth but decays eventually in the output port. Figure 6(d) represents the pure polariton gap phenomena and the field dies out rapidly through the first unit. Figure 6(b), however, accords to the EIT-like mini-passband caused by Bragg-polariton interference and the field somehow resonantly tunnels through the whole structure, yielding a nearly unity transmittance. From these panels it is apparent that Bragg gap arise from multi-scattering behavior associated with period characteristic, while polariton gap is related with the local resonant state. Mostly, they play their respective role in the interaction between light and matter when they are separated without overlap (see figure 6(d)). But when they approach each other, a sudden pass-band appears in the stop-band of polariton gap (see figure 6(b)).

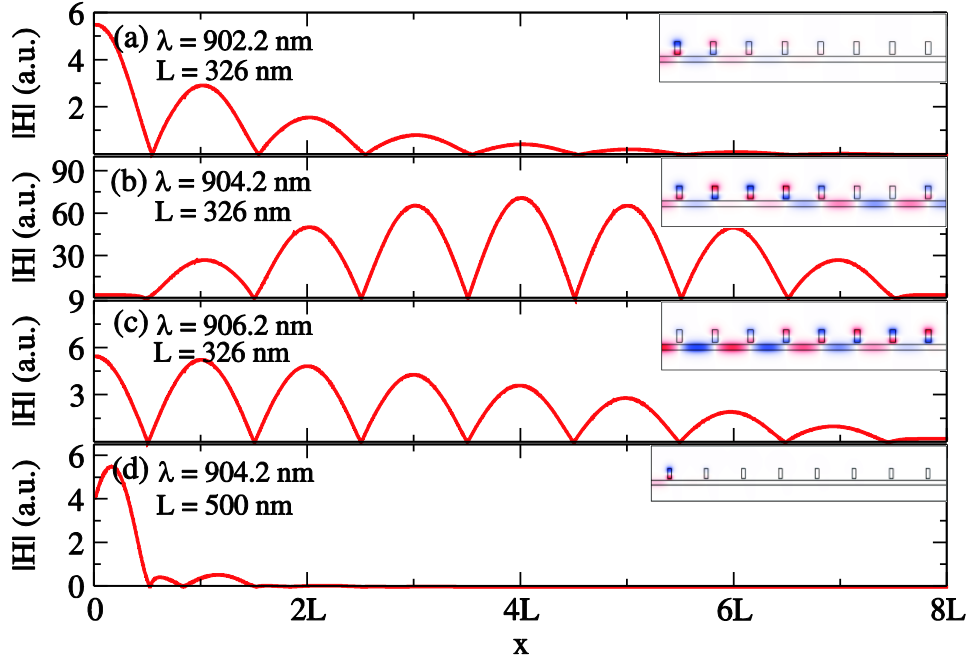

Figure 6. Normalized magnetic amplitude $|H_z|$ at the center line of the slot waveguide at different wavelength. Insets in each panel are the $\mathrm{Re}(H_z)$ distributions.

Finally, to justify the effectiveness of our approach, we consider another periodic structure as shown in figure 4(b) that has two resonators in each unit cell. In this case, the hybridized resonances shall yield two polariton gaps at $\lambda = 860$ nm and $\lambda = 936$ nm according to Eq. (10). Figure 7 shows the transmission of samples with one cell and 8 periods, for $L = 338$ nm and $L = 500$ nm, respectively. Interaction between Bragg resonance and polariton resonance takes place for period length of $L = 338$ nm, raising a sharp transparency window near $\lambda = 936.28$ nm. In the meantime, the polariton gap near $\lambda = 860$ nm simply persists without EIT-like transparency window. The mechanism is similar to the former case of one resonator side coupled to a waveguide. The difference is that two polariton resonances, arising from the hybrid modes of double resonators, interact with the tuned Bragg resonance. Similar to the former case, we have also get $(\gamma_e, \mu)$ pair (14.73 THz, 86.9 THz) in Eq. (9) from the unit cell result, and then get the theoretical spectrum by the TMM [labeled as TCMT-TMM]. Numerical results and the theoretical ones are also in good agreement.

A few comments are in order. Though we show results from both theoretical analysis and numerical simulation only for the cases of one and two local resonances. The method is applicable for more complex resonator-waveguide system. Nevertheless, it fails once the direct coupling between side-coupled resonators in nearby units is not negligible. Furthermore, it is possible to scale the design to work in frequency band that minimizes the metal Joule loss (e.g., in microwave or THz band). The resonator property could be dynamically tuned to achieve switching functionality. Yet to fully account the loss and possible gain in the system, a more general coupled mode theory shall be involved [27].

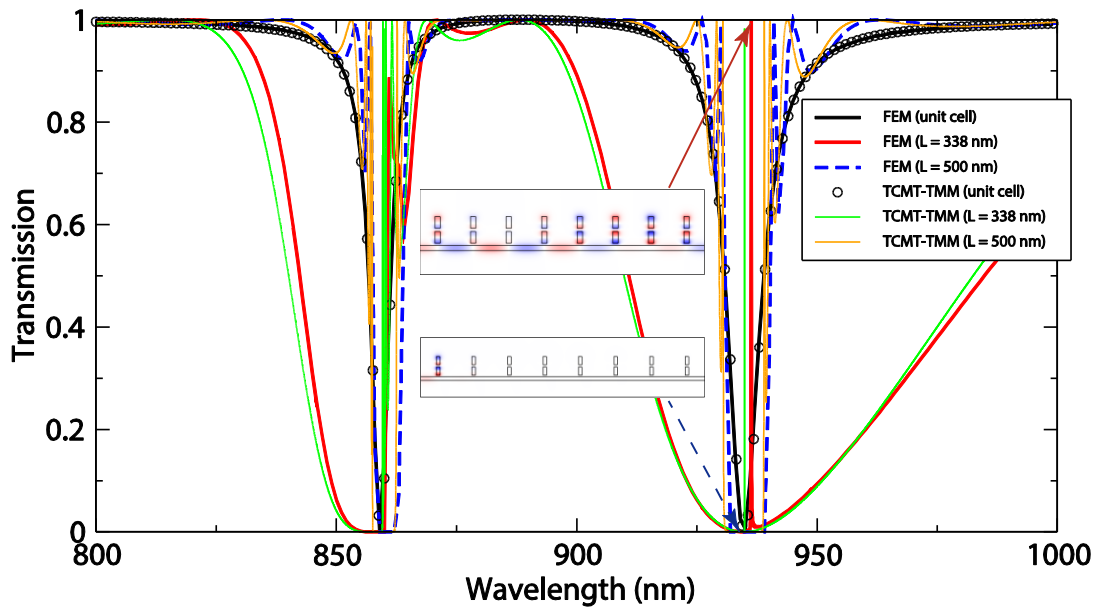

Figure 7. Full-wave numerical transmission spectrum of periodic structure in figure 4(b) for one unit cell (full black), multiple cells of periodic length $L = 338$ nm (full red) and $L = 500$ nm (dashed blue), and corresponding TCMT-TMM results for a unit cell (circle black), periodic length of 338 nm (full green) and 500 nm (full orange). The period number is 8 and absorption loss is neglected. . Insets are the numerical field patterns at $\lambda = 936.28$ nm for the EIT and non-EIT cases.

## 4. Summary

In conclusion, we have developed a theoretical scheme that allows an efficient tuning of the dispersion relation and transmission of resonator-waveguide coupled structures. We show that the interplay between two types of gaps—the Bragg gap and the polariton gap— can lead to multiple EIT-like phenomena. By combining the transfer matrix method and the temporal coupled mode theory, our scheme helps to investigate optical properties in cavity-associated structures and to design structure for specific guiding functionalities. By applying the theory to MDM waveguide system, we numerically show that with local resonances (single or double), the structure exhibits an extremely narrow transparency window via the interplay between the Bragg resonance and the polariton resonance.  These results would be useful in deep sub-wavelength photonic circuit design.

**Funding.** National Natural Science Foundation of China (Nos. 11274083 and 11004043), the NSF of Guangdong Province (No. 2015A030313748), and Shenzhen Municipal Science and Technology Plan (Nos. KQCX20120801093710373 and JCYJ201505133000065).